\documentclass[prl,twocolumn,floatfix,superscriptaddress,showpacs]{revtex4-1}
\usepackage{amsmath, amssymb}
\usepackage{graphicx,color}
\renewcommand{\section}[1]{{\par\it #1.---}\ignorespaces}
\begin{document}

\title{Detection of Majorana bound states by thermodynamically stable $4\pi$-periodic D.C.\, Josephson current}
\author{Zhan Cao}
\affiliation{Center for Interdisciplinary Studies $\&$ Key Laboratory for
Magnetism and Magnetic Materials of the MoE, Lanzhou University, Lanzhou 730000, China}
\author{Tie-Feng Fang}
\affiliation{Center for Interdisciplinary Studies $\&$ Key Laboratory for
Magnetism and Magnetic Materials of the MoE, Lanzhou University, Lanzhou 730000, China}
\author{Hong-Gang Luo}
\affiliation{Center for Interdisciplinary Studies $\&$ Key Laboratory for
Magnetism and Magnetic Materials of the MoE, Lanzhou University, Lanzhou 730000, China}
\affiliation{Beijing Computational Science Research Center, Beijing 100084, China}

\begin{abstract}
We propose a scheme to detect the Majorana bound states (MBSs) by a thermodynamically stable D.C.\, Josephson current with $4\pi$-periodicity in the superconducting phase difference, which is distinct from the previous A.C.\, $4\pi$-periodicity found in topological superconducting Josephson junctions. The scheme, consisting of a quantum dot coupled to two s-wave superconducting leads and a floating topological superconductor supporting two MBSs at its ends, only exploits the interplay of a local Zeeman field and the exotic helical and self-Hermitian properties of MBSs, without requiring the conservation of fermion parity and not relying on the zero-energy property of MBSs. Our D.C.\, $4\pi$-periodicity is thus robust against the overlap between the two MBSs and various system parameters, including the local Coulomb interaction, the tunneling asymmetry, and the width of superconducting gap, which facilitates experimentally detection of the MBSs.
\end{abstract}
\maketitle

\section{Introduction}\label{intr}
A Majorana fermion (MF) \cite{Majorana1937} is a fermion that is its own antiparticle. This exotic particle obeys non-Abelian statistics \cite{Leijnse2012} and can be manipulated by braid operators to realize quantum gates free of local decoherence \cite{Alicea2011,Beenakker2013}. Due to this promising application, the search and detection of MFs are currently under intensive studies. Whereas there is no experimental evidence for an elementary particle as a MF in nature until now, a number of theoretical predictions have pointed that quasi MFs can exist as low-energy excitations (i.e., MBSs) in condensed matter systems, such as the fractional quantum Hall systems \cite{Moore1991}, the vortex in a $p$-wave superconductor \cite{Kitaev2001,Ivanov2001}, two-dimensional topological insulators in proximity to a superconductivity source \cite{Fu2008}, and one-dimensional topological superconducting nanowires (TSNWs) \cite{Lutchyn2010,Oreg2010}. In particular, the TSNW, which possesses two MBSs each seated at its two ends, can be easily realized by setting a semiconductor nanowire (e.g., InAs/InSb) with strong spin-orbit interaction coupled proximally to an s-wave superconductor and exposed to an external magnetic field \cite{Mourik2012,Deng2012,Das2012,Churchill2013,Lee2014}.

Most of the schemes to detect MBSs have focused on the electronic transport properties, which can be significantly altered by the MBSs and exhibit signatures for detection, for instance, the zero-bias peak in the conductance \cite{Mourik2012,Deng2012,Das2012,Churchill2013,Sau2010,Liu2012,Prada2012,Pientka2012}, the phase transition signaled by quantized shot noise and a period doubling of the magnetoconductance oscillations \cite{Akhmerov2011}, resonant Andreev reflections with parity-dependent $0$ or $2e^2/h$ conductance \cite{Law2009}, and various nonlocal tunneling phenomena \cite{Nilsson2008,Fu2010,Benjamin2010}. Moreover, one of the remarkable signatures of MBSs is the fractional Josephson effect predicted and probed in a topological superconducting (TS) Josephson junction consisting of two TSNWs \cite{Kitaev2001,Lutchyn2010,Oreg2010,Jiang2011,Badiane2011,Jose2012,Dominguez2012,Pikulin2012,Rokhinson2012,Badiane2013,Pekker2013}, where the supercurrent has been predicted to exhibit a $4\pi$-periodicity in the superconducting phase difference, instead of $2\pi$ as in usual Josephson junctions, due to the fact that the supercurrent is carried by a single Majorana particle rather than the Copper pairs. However, such a $4\pi$-periodicity, requiring the parity conservation and adiabatically tuning the phase, is very difficult to observe in the usual D.C. measurements where the inelastic relaxation processes are inevitable and thus the parity conservation is often violated (the so-called quasiparticle poisoning) \cite{Jiang2011,Badiane2011,Pikulin2012}. Therefore, one must resort to the complicated and indirect A.C. supercurrent measurement in the condition that the time period of Josephson oscillations is shorter than the inelastic relaxation time \cite{Jiang2011,Jose2012,Pikulin2012,Dominguez2012}. In this case, the $4\pi$-periodicity manifest itself as a doubling of the first Shapiro steps in the supercurrent-voltage characteristics and has been indeed observed experimentally \cite{Rokhinson2012}.

\begin{figure}[tbp]
\begin{center}
\includegraphics[width=0.9\columnwidth]{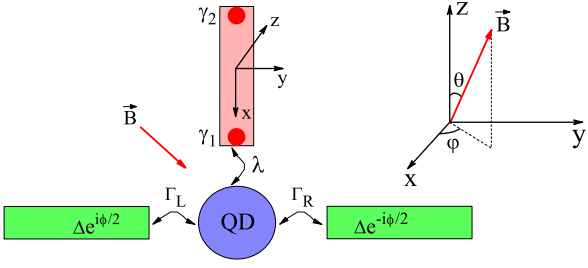}
\caption{(Color online) Schematic view of our system: a QD coupled to the left and right superconducting leads and also to one end of a floating TSNW. The Rashba-type spin-orbit coupling in the nanowire is along the $y$ direction and the magnetic field needed to drive the nanowire to topological nontrivial phase is along the $x$ direction. Two MBSs $\gamma_{1,2}$ emerge at the two ends of the wire. Only the nearest MBS $\gamma_1$ directly couples to the QD with strength $\lambda$. An independent magnetic filed $\vec B$ with azimuth angle $(\theta, \varphi)$ is applied on the dot.}\label{fig1}
\end{center}
\end{figure}

In this paper, we propose a scheme to detect the MBSs by the thermodynamically stable $4\pi$-periodic D.C.\,Josephson current without invoking the complication of A.C. measurements. The setup is shown in Fig.\,\ref{fig1}, a quantum dot (QD) is inserted between the left and right s-wave superconducting leads and is also side-coupled to one end of a floating TSNW. A local magnetic field is applied on the dot to induce a Zeeman splitting. For such a system, the interplay of the local Zeeman field and the exotic helical and self-Hermitian properties of MBSs gives rise to a $4\pi$-periodic D.C.\,Josephson current through the QD, which does not require the parity conservation and is not influenced by inelastic relaxation processes. Since this D.C.\,$4\pi$-periodicity is robust against the length of the TSNW, the asymmetry of tunneling rates, the local Coulomb interaction, as well as the width of superconducting gap, we suggest that by directly measuring this feature the characteristics of MBSs in the TSNW could be identified. Similar side-coupled configurations but with normal leads recently have been also proposed to probe the MBSs \cite{Liu2011,Cao2012,Lee2013,Vernek2014,Lopez2014,Gong2014}.

\section{Model}\label{model}
Our system of Fig.\,\ref{fig1} can be described by the following Hamiltonian:
\begin{equation}
H = H_L+H_R+H_{C}+H_{T}, \label{eq1}
\end{equation}
where $H_{l}$ ($l=L,R$) is the standard BCS Hamiltonian for the left (L) and right (R) s-wave superconducting leads with an energy gap $\Delta$,
\begin{equation}
H_{l}=\sum_{k,\sigma}\varepsilon_{k}c_{kl\sigma}^{\dag}c_{kl\sigma}-\sum_{k}(\Delta e^{i(\delta_{lL}-\delta_{lR})\phi/2}c_{kl\uparrow}^{\dag}c_{-kl\downarrow}^{\dag}+\textrm{H.c.}), \label{eq2}
\end{equation}
Experimentally, the superconducting phase difference $\phi$ across the junction can be tuned by a magnetic flux passing through a superconducting ring formed by the two leads \cite{Sadovskyy2012}. $c^\dag_{kl\sigma}$ ($c_{kl\sigma}$) is the creation (annihilation) operators of the electron in the $l$ lead with energy $\varepsilon_{k}$. The chemical potentials of the left and right leads are set as $\mu=0$.

The third term in Eq.\,(\ref{eq1}) $H_C$ models the central region of a quantum dot side-coupled to one end of the TSNW,
\begin{eqnarray}
&&H_{C}=\sum_{\sigma}\varepsilon_{d} d_{\sigma}^{\dag}d_{\sigma}+\vec B\cdot \vec S+Un_{d\uparrow}n_{d\downarrow}+i\varepsilon_{M}\gamma_{1}\gamma_{2}\notag\\
&&\hspace{0.8cm}+\sqrt{2}\lambda( d_\uparrow-d_\uparrow^{\dag})\gamma_{1}, \label{eq3}
\end{eqnarray}
where $d_\sigma^{\dag}$ ($d_\sigma$) represents the creation (annihilation) operators of the dot electron with spin $\sigma$, and $n_\sigma\equiv d_\sigma^{\dag}d_\sigma$. The dot level is degenerate as $\varepsilon_d$ since the magnetic field required to drive the nanowire to topological nontrivial phase is usually not so strong to induce the Zeeman splitting due to the small $g$ factor in QD \cite{Lee2013,note3}. To induce a significant QD Zeeman splitting, we apply an independent strong magnetic filed only on the QD $\vec B=(B\sin\theta\cos\varphi,B\sin\theta\sin\varphi,B\cos\theta)$, as described by the second term in Eq.\,(\ref{eq3}). Under the basis of QD electron, the three components of the dot spin operator $S$ can be expressed explicitly as $S_x=\frac{1}{2}(d^\dag_\uparrow d_\downarrow+d^\dag_\downarrow d_\uparrow)$, $S_y=\frac{1}{2i}(d^\dag_\uparrow d_\downarrow-d^\dag_\downarrow d_\uparrow)$, and $S_z=\frac{1}{2}(d^\dag_\uparrow d_\uparrow-d^\dag_\downarrow d_\downarrow)$. Therefore, the first two terms of Eq.\,(\ref{eq3}) can be collected as $\sum_\sigma\varepsilon_{d\sigma}d_{\sigma}^{\dag}d_{\sigma}+\mathcal{F}(e^{-i\varphi}d^\dag_\uparrow d_\downarrow+ H.c.)$ with $\varepsilon_{d\sigma}=\varepsilon_d+\frac{1}{2}\sigma B\cos\theta$ and $\mathcal{F}=\frac{1}{2}B\sin\theta$. The local Coulomb interaction on the dot is denoted by $U$. The topological phase in the nanowire is described by a low-energy effective model in which the two zero-energy MBSs are represented by the Majorana operators $\gamma_i$ ($i=1,2$), obeying the Clifford algebra $\{\gamma_i,\gamma_j\}=\delta_{ij}$ and $\gamma_i=\gamma_i^\dag$. Moreover, for a finite-length nanowire, the two MBSs couple to each other by an nonzero overlap energy $\varepsilon_{M}$. Due to the helical property of the Majorana end states, only the nearest MBS couples to one of the spin orientations (say spin-up) of dot electrons with real coupling strength $\lambda$ \cite{Lee2013,Leijnse2011,Golub2011,note2}, resulting in a polarized hybridization.

The last term in Eq.\,(\ref{eq1}) describes the tunneling between the QD and the superconducting leads, which reads
\begin{equation}
H_{T}=\sum_{k,l\sigma}(t_lc_{kl\sigma}^{\dag}d_\sigma+\textrm{H.c.}), \label{eq4}
\end{equation}
with $t_l$ being the tunneling matrix elements. An electron and/or hole transfer between the dot and the leads is described by an effective tunneling rate $\Gamma_l$, which in the wide-band approximation takes the form $\Gamma_l=\pi|t_l|^2\rho$, where $\rho$ is the density of normal states in superconducting electrode. The total tunneling rate $\Gamma=\Gamma_L+\Gamma_R=1$ is used as an energy unit throughout this paper.

In order to achieve an analytical insight for the origin of our D.C.\, $4\pi$-periodicity, we start with the so-called atomic limit $\Delta\rightarrow\infty$. In this regime, only constant off-diagonal self-energies on the QD survive \cite{Sadovskyy2012,Hewson2007,Karrasch2008,Meng2009}, which give rise to a local pairing between the dot electrons. We thus arrive at an effective Hamiltonian \cite{Meng2009}
\begin{eqnarray}
&&H_\textrm{eff}=\sum_{\sigma}\varepsilon_{d\sigma}d_{\sigma}^{\dag}d_{\sigma}+\mathcal{F}(e^{-i\varphi}d^\dag_\uparrow d_\downarrow+ H.c.)+Un_{d\uparrow}n_{d\downarrow} \notag\\
&&-(\widetilde{\Gamma}d_{\uparrow}^{\dag}d_{\downarrow}^{\dag}+H.c.)+\varepsilon_{M}f^\dag f+\lambda( d_\uparrow-d_\uparrow^{\dag})(f+f^\dag), \label{eq5}
\end{eqnarray}
with $\widetilde\Gamma=\sum_l\Gamma_l e^{i(\delta_{lL}-\delta_{lR})\phi/2}$. Here the Majorana operators $\gamma_{1,2}$ have been replaced with regular fermion operator $f$ through the transformation $\gamma_1=\frac{f+f^\dag}{\sqrt{2}}$ and $\gamma_2=\frac{f-f^\dag}{i\sqrt{2}}$.

Since the dimension of the Hilbert space of this effective Hamiltonian is limited ($D=8$), exact results are available at least numerically. Moreover, the parity symmetry of the system, $[H_\textrm{eff},\ (-1)^{\hat N}]=0$, with $\hat{N}=n_\uparrow+n_\downarrow+f^\dagger f$, allows to classify the eigenstates according to the even ($e$) and odd ($o$) parity:
\begin{eqnarray}
|\Psi_m^e\rangle=a_m^e|0,0\rangle+b_m^e|\uparrow\downarrow,0\rangle+c_m^e|\uparrow,1\rangle+d_m^e|\downarrow,1\rangle,\label{eq6}\\
|\Psi_m^o\rangle=a_m^o|0,1\rangle+b_m^o|\uparrow\downarrow,1\rangle+c_m^o|\uparrow,0\rangle+d_m^o|\downarrow,0\rangle,\label{eq7}
\end{eqnarray}
where $m=1,2,3,4$ and the basis $|\mu,\nu\rangle=|\mu\rangle\otimes|\nu\rangle$ is a tensor product of the dot state $|\mu\rangle=|0\rangle,|\uparrow\rangle,|\downarrow\rangle,|\uparrow\downarrow\rangle$ and the Majorana state $|\nu\rangle=|0\rangle,|1\rangle$. For the special case of $\varepsilon_M=U=0$, we obtain simple analytical results for the eigenenergies which are degenerate for the two subspaces, $E_{1,2}^e=E_{1,2}^o=\varepsilon_d\mp a_2$, $E_{3,4}^e=E_{3,4}^o=\varepsilon_d\mp a_1$, with $a_{1,2}=\frac{1}{\sqrt{2}}\sqrt{\Lambda+\frac{1}{4}B^{2}\mp\sqrt{\left(  \Lambda-\frac{1}{4}B^{2}\right)  ^{2}+\Omega+\Xi}}$, $\Lambda=\vert \widetilde{\Gamma}\vert ^{2}+2\lambda^{2}+\varepsilon_{d}^{2}$, $\Omega=4\lambda^{2}\left( \frac{1}{2}B^{2}-\lambda^{2}-\varepsilon_{d}B\cos\theta\right)$, and $\Xi =4\lambda^{2}B\sin\theta\operatorname{Re}[\widetilde{\Gamma}e^{-i\varphi}]$. Note that $|\widetilde\Gamma|^2=\Gamma_L^2+\Gamma_R^2+2\Gamma_L\Gamma_R\cos\phi$ and $\operatorname{Re}[\widetilde{\Gamma}e^{-i\varphi}]$ are functions with $2\pi$ and $4\pi$-periodicity with respect to the phase difference $\phi$, respectively. As an immediate consequence, all energies will vary with the phase difference $\phi$ in distinct $4\pi$-periodicity provided that $\Xi\neq 0$. This is one of our central results. Physically, this $4\pi$-periodicity feature is attributed to the interplay of the exotic helical and self-Hermitian properties of MBSs with the Zeeman splitting on the dot level. We emphasize that the predicted $4\pi$-periodicity is not restricted to the special parameters used here. It also shows up for general parameters $\varepsilon_M,\,U\neq0$ and even for finite gap width $\Delta$ (as we numerically verify below).

\begin{figure}[tbp]
\begin{center}
\includegraphics[width=\columnwidth]{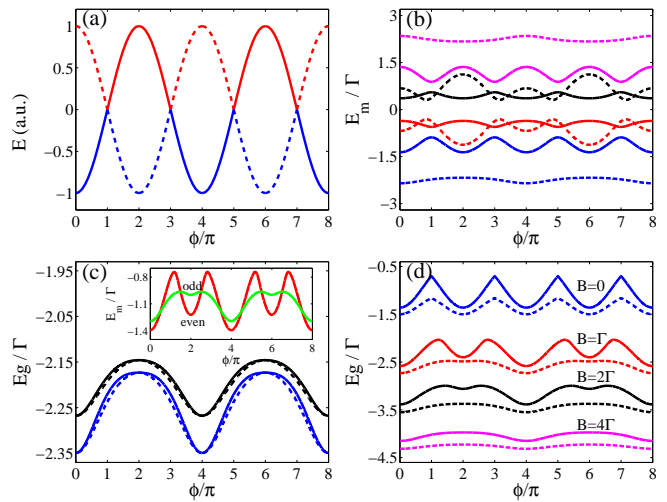}
\caption{(Color online) (a) Energies of previous TS Josephson junction in odd (solid line) and even (dashed line) parity. The parity of ground state (blue line) changes with the phase difference $\phi$ alternatively. (b)-(d) Eigenenergies of our system in the large gap limit. (b) Full energy spectrum for $\Gamma_L/\Gamma_R=2$, $B=0$ (solid lines), and $B=4\Gamma$ (dashed lines). The dashed lines are $4\pi$-periodic with respect to $\phi$. (c) Effect of asymmetric tunneling rates ($\Gamma_L/\Gamma_R=1$, solid lines; $\Gamma_L/\Gamma_R=4$, dashed lines) and the overlap ($\varepsilon_M=0$, blue lines; $\varepsilon_M=0.5$, black lines) between the two MBSs on the ground energy. Inset: two minimal eigenenergies in even and odd parity subspaces for $B=1.4\Gamma$ and $\varepsilon_M=0.5\Gamma$. (d) Ground energy of noninteracting $U=0$ (solid lines) and interacting $U=2\Gamma$ (dashed lines) QD for different Zeeman splitting. The curves in (d) are offset for clarity. Other parameters are $\Gamma_L/\Gamma_R=1$, $B=4\Gamma$, $U=0$, $\varepsilon_d=-U/2$, $\varepsilon_M=0$, $\lambda=0.7\Gamma$, and $(\theta,\,\varphi)=(\pi/3,\,0)$ unless specified.}\label{fig2}
\end{center}
\end{figure}
\section{Ground energy}\label{energy}
The energy spectrum of the previous TS Josephson junction \cite{Jiang2011,Badiane2011} is $E_A(\phi)\sim -(n_A-1/2)\Delta\cos\frac{\phi}{2}$ with $n_A$ the occupation of the Andreev bound state. This is a $4\pi$-periodic function for certain parity, as shown in Fig.\,\ref{fig2}(a). However, the two levels cross periodically as the phase difference varies, leading to a ground energy of $2\pi$-periodicity. In the D.C.\, measurement at thermal equilibrium, especially for $T \ll |E_A(\phi)|$, the inevitable inelastic processes will always violate the parity conservation and the system will immediately relax to the ground state whose parity depends on the phase difference $\phi$. Therefore, one actually obtains a $2\pi$-periodic D.C.\,Josephson current.

This is not the case for our scheme. The energy spectrum of our system is given in Fig.\,\ref{fig2}(b)-(d). It is shown that the periodicity in the phase difference $\phi$ of all energy levels change from $2\pi$ to $4\pi$ when the coupling to MBSs $\lambda$ and the local Zeeman field $B$ coexist [Fig.\,\ref{fig2}(b)], due to the interplay of the dot level splitting, the polarized dot-Majorana coupling, and the self-Hermitian property of such end MBSs. Moreover, our $4\pi$-periodicity feature is rather robust against the asymmetric tunneling rates as well as the overlap between the two MBSs [Fig.\,\ref{fig2}(c)]. The latter is for the reason that the prediction does not rely on the zero-energy property of MBSs. This is in advantage over the scheme mentioned in Ref.\,[\onlinecite{Liu2011}], where the Majorana signature of half-quantum resonant linear conductance is fragile even for a very small $\varepsilon_M$. Note that for some special choices of parameters, our energy levels as a function of $\phi$ also show crosses [inset of Fig.\,\ref{fig2}(c)]. However, the resulting ground energy are still $4\pi$-periodic even though the parity is no longer conserving, being distinct from the previous results of Fig.\,\ref{fig2}(a). Therefore, our $4\pi$-periodicity can indeed survive from inelastic relaxation processes in the D.C.\, measurement. Finally, it is also shown that the $4\pi$-periodicity feature becomes more evident (meaning more dominant $4\pi$ component and less $2\pi$ or other components in the Fourier space) as the local Coulomb interaction turns on and the Zeeman field increases [Fig.\,\ref{fig2}(d)].

\section{D.C.\,Josephson current}\label{current}
The above remarkable $4\pi$-periodicity in the energy spectrum is directly measurable by examining the D.C. Josephson current. Within the framework of Green's functions (GFs) \cite{Haug2008,Sun2000,Osawa2008}, the thermal equilibrium D.C.\,Josephson current through our system (1) is formulated as
\begin{equation}
J=\frac{2e}{h}\frac{4\Gamma_L\Gamma_R}{\Gamma}\int d\omega f(\omega)j(\omega), \label{eq12}
\end{equation}
with $f(\omega)$ the Fermi distribution function and the supercurrent density is \cite{Note}
\begin{equation}
j(\omega)=\Delta\sin\frac{\phi}{2}\operatorname{Im}\{\beta(\omega) [\langle\langle d_{\uparrow},d_{\downarrow}\rangle \rangle^{r}_\omega+\langle \langle d_{\downarrow}^\dag,d_{\uparrow}^\dag\rangle\rangle^{r}_\omega]\}, \label{eq13}
\end{equation}
where $\beta(\omega)=\frac{-1}{\sqrt{\Delta^2-(\omega+i 0^+)^2}}$ and $\langle\langle\cdot,\cdot\rangle\rangle^r_\omega$ is the retarded GF. In the large gap limit $\Delta\rightarrow\infty$, the supercurrent is readily reduced to \cite{Note}
\begin{equation}
J=\frac{2e}{h}\frac{8\pi\Gamma_{L}\Gamma_{R}}{\Gamma}\sin\frac{\phi}{2}\operatorname{Re}\langle d_{\downarrow}d_{\uparrow}\rangle, \label{eq14}
\end{equation}
where $\langle d_{\downarrow}d_{\uparrow}\rangle$ is thermal average over all the eigenstates of Eqs.\,(\ref{eq6}) and (\ref{eq7}). Note that at zero temperature only products between the weights of empty and doubly-occupied dot level in the ground state, $\sim a_m^*b_m$, contribute to the average. To take account of the effect of finite gap width on the $4\pi$-periodicity, we have also exactly solved the original Hamiltonian (1) in the noninteracting ($U=0$) case. The GFs needed in Eq.\,(\ref{eq13}) for calculating the supercurrent are then obtained through the Dyson equation $G^{-1}(z)=G^{-1}_0(z)-\Sigma(z)$ with $G_0(z)=(zI-H_C)^{-1}$ being the unperturbated GFs of the central region and $\Sigma(z)$ the self-energy resulting from the coupling to the superconducting leads \cite{Note}.

\begin{figure}[tbc]
\begin{center}
\includegraphics[width=\columnwidth]{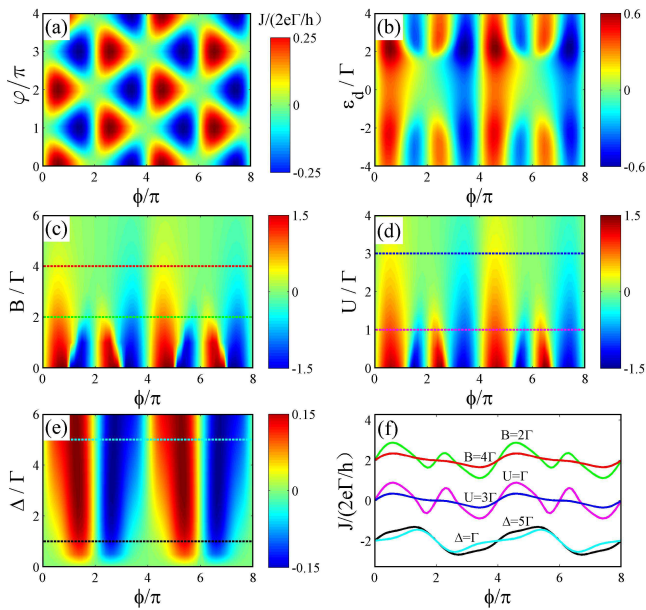}
\caption{(Color online) (a) Josephson current as a function of $\phi$ and $\varphi$. (b)-(e) Effect of the dot level (b), Zeeman splitting (c), Coulomb interaction [(d), $B=\Gamma$], and finite gap width (e) on the $4\pi$-periodicity. (f) Lines cut at (c), (d), and (e) with offset $2$, $0$, and $-2$, respectively. The two lines cut at (e) are also multiplied by $4$ for clarity. The temperature is set as $T=0$ and other parameters are the same as those in Fig.\ref{fig2} unless specified.}\label{fig3}
\end{center}
\end{figure}

We present results for the Josephson current in Fig.\,\ref{fig3}. At finite magnetic field and dot-Majorana coupling, the Josephson current indeed exhibits the $4\pi$-periodicity with respect to the phase difference $\phi$ across the junction, as shown in Fig.\,\ref{fig3}(a) (where the $2\pi$-periodicity with the azimuth angle $\varphi$ of the magnetic field is trivial). By sweeping the dot level over the particle-hole(p-h) symmetric point ($\varepsilon_d=0$ in the $U=0$ case), we find that the $4\pi$-periodicity is more distinct at the p-h symmetric point, and the supercurrent is asymmetric on the two sides of this point due to the violated space rotation symmetry by the magnetic filed [Fig.\,\ref{fig3}(b)]. Interestingly, the effects of magnetic field and Coulomb interaction on the periodicity of the current are quite similar: dominant $4\pi$-periodicity shows up in the supercurrent with decreased amplitude when the magnetic field or the Coulomb interaction increases [Figs.\,\ref{fig3}(c), (d), and (f)]. The underlying physics are also similar. Note that at the p-h symmetric point, the weight $b^{e(o)}$ of a doubly-occupied dot level in the ground state [of the form Eqs.\,(\ref{eq6}) or (\ref{eq7})] can be reduced either by the magnetic field or the Coulomb interaction. Since this weight contributes effectively to the supercurrent as mentioned before, the current amplitude is certainly suppressed. For a noninteracting QD at the p-h symmetric point without coupling to the MBSs, the suppercurrent is of course $2\pi$-periodic. When the interaction $U$ turns on, this $2\pi$-periodic current is found \cite{Hewson2007} to vanish immediately in the large gap limit. One thus naturally expect that the Coulomb interaction can also largely suppress the $2\pi$-periodic component of the supercurrent which is less related to the Majorana physics, even in the presence of coupling to MBSs. This explains why the $4\pi$-periodicity, which inherits most of Majorana characteristics, becomes more distinct as $U$ increases.

For finite gap width, the supercurrent is carried by the continuous and discrete (Andreev bound states) spectrum of the QD. Both contribute with opposite sign to the total current. Majorana physics is included in the discrete spectrum via direct coupling to the QD. As the gap width increases, the contribution from the discrete spectrum prevails in the transport, leading to the increased current amplitude and evident $4\pi$-periodicity [Fig.\,\ref{fig3}(e) and (f)].

\section{Conclusions}\label{sum}
We have predicted a $4\pi$-periodic D.C.\,Josephson current due to the interplay of the Zeeman field and the exotic helical and self-Hermitian properties of MBSs in the system of a QD coupled to two s-wave superconductors and one TNSW. As a signature of Majorana Fermions in supercurrent, we emphasize that this $4\pi$-periodicity is thermodynamically stable over a wide range of system parameters and thus facilitates experimentally detection of MBSs by D.C. measurements, looking more feasible than the previous schemes using A.C. Josephson effect \cite{Jiang2011,Pikulin2012,Dominguez2012,Rokhinson2012}. We hope that our prediction will indeed motivate experimental effort in this direction.

\section{Acknowledgments}
This work is supported partly by NSFC, PCSIRT (Grant No. IRT1251), and the national program for basic research of China.


\begin{thebibliography}{99}
\bibitem{note3} Note that even though the magnetic filed driving the nanowire into the topological phase already induces a significant Zeeman splitting on the dot, our $4\pi$-periodicity still survives. In this case, since the dot-MBS coupling and the Zeeman splitting both depend on the magnetic field, another local magneitc field applied only on the dot is still needed to independently control the Zeeman splitting and the dot-MBS coupling.
\bibitem{note2} Generally, $\lambda$ is a complex number with a phase factor from the TSNW. This phase can be absorbed into $d_\uparrow$ and $c_{kl\uparrow}$ operators by proper gauge transformations, thereby shifting the azimuth angle $\varphi$ of $\vec B$ and the phase of the left and right superconducting leads by the same magnitude. Such a shifting does not change any intrinsic phyiscs of the predicted $4\pi$-periodicity. We thus treat $\lambda$ as a real number.
\end{thebibliography}
\end{document}


\title{Supplemental materials}
\author{Zhan Cao}
\affiliation{Center for Interdisciplinary Studies $\&$ Key Laboratory for
Magnetism and Magnetic Materials of the MoE, Lanzhou University, Lanzhou 730000, China}
\author{Tie-Feng Fang}
\affiliation{Center for Interdisciplinary Studies $\&$ Key Laboratory for
Magnetism and Magnetic Materials of the MoE, Lanzhou University, Lanzhou 730000, China}
\author{Hong-Gang Luo}
\affiliation{Center for Interdisciplinary Studies $\&$ Key Laboratory for
Magnetism and Magnetic Materials of the MoE, Lanzhou University, Lanzhou 730000, China}
\affiliation{Beijing Computational Science Research Center, Beijing 100084, China}
\maketitle
\section{FORMULA OF JOSEPHSON CURRENT}
In this section, we present the details for deriving the formula of thermal equilibrium D.C.\, Josephson current. The supercurrent from lead $l$ to the central region can be calculated from the time evolution of the occupation number operator \cite{Haug2008}
\begin{eqnarray}
&&J_{l}=e\langle\dot N_l\rangle \notag\\
&&\hspace{0.4cm}=\frac{e}{i\hbar}\langle [  \sum_{k\sigma}n_{kl\sigma}  ,H  ]  \rangle \notag\\
&&\hspace{0.4cm}=-\frac{e}{i\hbar}\sum_{k\sigma}t_{l}[  \langle d_{\sigma}^{\dag}c_{kl\sigma}\rangle -\langle c_{kl\sigma}^{\dag}d_{\sigma}\rangle ] \notag\\
&&\hspace{0.4cm}=-\frac{e}{i\hbar}\sum_{k}t_{l}[  (  \langle d_{\uparrow}^{\dag}c_{kl\uparrow}\rangle -\langle c_{kl\uparrow}^{\dag}d_{\uparrow}\rangle )  -(  \langle c_{-kl\downarrow}^{\dag}d_{\downarrow}\rangle -\langle d_{\downarrow}^{\dag}c_{-kl\downarrow}\rangle )  ] \notag\\
&&\hspace{0.4cm}=-\frac{2e}{\hbar}\sum_{k}t_{l}\operatorname{Im}[  \langle d_{\uparrow}^{\dag}c_{kl\uparrow}\rangle +\langle d_{\downarrow}c_{-kl\downarrow}^{\dag}\rangle ]. \label{A1}
\end{eqnarray}
In the equilibrium case, the fluctuation-dissipation theorem (FDT) relates the expectation value to the retarded and advanced Green functions (GFs) as \cite{Haug2008}
\begin{equation}
\langle AB\rangle=\frac{i}{2\pi}\int d\omega f(\omega)(\langle \langle B,A \rangle \rangle ^r_\omega - \langle \langle B,A \rangle \rangle ^a_\omega), \label{A2}
\end{equation}
where $f(\omega)=1/(e^{\omega/k_B T}+1)$ is the Fermi distribution function. The retarded and advanced GFs are defined as $\langle \langle B(t),A(t') \rangle \rangle^{r,a}=\mp i\theta(\pm t\mp t') \langle \{B(t),A(t')\}\rangle$, and $\langle \langle B,A \rangle \rangle ^{r,a}_\omega$ are the Fourier-transformed GFs, correspondingly. Using the equation of motion method \cite{Haug2008}, the retarded (advanced) GFs $\langle \langle c_{kl\uparrow},d^\dag_\uparrow \rangle \rangle ^{r,a}_\omega$ and $\langle \langle c_{-kl\downarrow}^\dag,d_\downarrow \rangle \rangle ^{r,a}_\omega$ are readily obtained as
\begin{eqnarray}
&&\langle \langle c_{kl\uparrow},d_{\uparrow}^{+}\rangle\rangle^{r,a}=t_{l}\frac{\Delta_{l}\langle\langle d_{\downarrow}^{+},d_{\uparrow}^{+}\rangle \rangle^{r,a} +\left(  z_{\pm}+\varepsilon_{kl}\right) \langle \langle d_{\uparrow},d_{\uparrow}^{+}\rangle\rangle^{r,a} }{\left(z_{\pm}+\varepsilon_{kl}\right)  \left(z_{\pm}-\varepsilon_{kl}\right)  -\left\vert \Delta\right\vert ^{2}}, \label{A3}\\
&&\langle \langle c_{-kl\downarrow}^{+},d_{\downarrow}\rangle\rangle^{r,a}=-t_{l}\frac{\Delta_{l}^{\ast}\langle \langle d_{\uparrow},d_{\downarrow}\rangle \rangle^{r,a} +\left(  z_{\pm}-\varepsilon_{kl}\right) \langle \langle d_{\downarrow}^{+},d_{\downarrow}\rangle\rangle^{r,a} }{\left(  z_{\pm}+\varepsilon_{kl}\right)  \left(z_{\pm}-\varepsilon_{kl}\right)  -\left\vert \Delta\right\vert ^{2}},\label{A4}
\end{eqnarray}
where $\Delta_l=\Delta e^{i(\delta_{lL}-\delta_{lR})\phi/2}$ and $z_{\pm}=\omega\pm i 0^+$.
Combining above Eqs. the Josephson current reads
\begin{equation}
J_{l}=-\frac{4e}{h}\int d\omega f(\omega)\Gamma_l\operatorname{Re}\{\beta(\omega)  [\Delta_{l}\langle\langle d_{\downarrow}^\dag,d_{\uparrow}^\dag\rangle\rangle^{r}_\omega-\Delta_{l}^{\ast}\langle \langle d_{\uparrow},d_{\downarrow}\rangle \rangle_\omega ^{r}]  \}, \label{A5}
\end{equation}
where $\Gamma_l=\pi\rho t_l^2$ and $\beta(\omega)=\frac{1}{\pi}\int d\varepsilon\frac{1}{(\omega+i 0^++\varepsilon)(\omega+i 0^+-\varepsilon)- \Delta^{2}}=-\frac{1}{\sqrt{\Delta^2-(\omega+i 0^+)^2}}$. In steady state, the current will be uniform, so that $J=J_L=-J_R$. The current formula can be simplified by reexpressing $J=\frac{\Gamma_R}{\Gamma_L+\Gamma_R}J_{L}-\frac{\Gamma_L}{\Gamma_L+\Gamma_R}J_{R}$, which gives, using Eq. (\ref{A3}),
\begin{equation}
J=\frac{2e}{h}\frac{4\Gamma_L\Gamma_R}{\Gamma}\int d\omega f(\omega)j(\omega), \label{A6}
\end{equation}
with $\Gamma=\Gamma_L+\Gamma_R$, and the supercurrent density is
\begin{equation}
j(\omega)=\Delta\sin\frac{\phi}{2}\operatorname{Im}\{ \beta(\omega) [\langle\langle d_{\uparrow},d_{\downarrow}\rangle \rangle^{r}_\omega +\langle \langle d_{\downarrow}^\dag,d_{\uparrow}^\dag\rangle\rangle^{r}_\omega ]\}. \label{A7}
\end{equation}
Furthermore, in the atomic limit $\Delta\rightarrow\infty$, one can see that $\Delta\beta(\omega)\rightarrow-1$, thus
\begin{equation}
j(\omega)=-\sin\frac{\phi}{2}\operatorname{Im}[  \langle \langle d_{\uparrow},d_{\downarrow}\rangle \rangle_\omega ^{r}+\langle\langle d_{\downarrow}^{\dag},d_{\uparrow}^{\dag}\rangle\rangle_\omega^{r}]. \label{A8}
\end{equation}
Note that
\begin{eqnarray}
&&\int d\omega f(\omega)  \operatorname{Im}[  \langle\langle d_{\uparrow},d_{\downarrow}\rangle \rangle_\omega^{r}+\langle \langle d_{\downarrow}^{\dag},d_{\uparrow}^{\dag}
\rangle \rangle_\omega ^{r}]  \notag\\
&&\hspace{-0.4cm} =\int d\omega f(\omega)  \frac{1}{2i}[  \langle\langle d_{\uparrow},d_{\downarrow}\rangle \rangle_\omega^{r}-(  \langle \langle d_{\uparrow},d_{\downarrow}\rangle \rangle_\omega ^{r})  ^{\ast}+\langle \langle d_{\downarrow}^{\dag},d_{\uparrow}^{\dag}\rangle \rangle_\omega ^{r}-(
\langle \langle d_{\downarrow}^{\dag},d_{\uparrow}^{\dag}\rangle\rangle_\omega ^{r})  ^{\ast}]  \notag\\
&&\hspace{-0.4cm}=\int  d\omega f(\omega)  \frac{1}{2i}[  \langle\langle d_{\uparrow},d_{\downarrow}\rangle \rangle_\omega^{r}-\langle \langle d_{\downarrow}^{\dag},d_{\uparrow}^{\dag}\rangle \rangle_\omega ^{a}+\langle \langle d_{\downarrow}^{\dag},d_{\uparrow}^{\dag}\rangle \rangle_\omega ^{r}-\langle
\langle d_{\uparrow},d_{\downarrow}\rangle \rangle_\omega^{a}]  \notag\\
&&\hspace{-0.4cm}=-\frac{i}{2}\int d\omega f(\omega)[  \langle\langle d_{\uparrow},d_{\downarrow}\rangle \rangle_\omega^{r}-\langle \langle d_{\uparrow},d_{\downarrow}\rangle\rangle_\omega ^{a}+\langle \langle d_{\downarrow}^{\dag},d_{\uparrow}^{\dag}\rangle \rangle_\omega ^{r}-\langle \langle d_{\downarrow}^{\dag},d_{\uparrow}^{\dag}\rangle \rangle_\omega ^{a}] \notag\\
&&\hspace{-0.4cm}=-\pi(\langle d_\downarrow d_\uparrow\rangle +\langle d_\uparrow^\dag d_\downarrow^\dag\rangle)\notag\\
&&\hspace{-0.4cm}=-2\pi\operatorname{Re}\langle d_{\downarrow}d_{\uparrow}\rangle, \label{A9}
\end{eqnarray}
where we have employed the FDT Eq. (\ref{A2}) and the relation $\langle\langle A, B \rangle\rangle^r_\omega=(\langle\langle B^\dag, A^\dag \rangle\rangle^a_\omega)^*$. Thus
\begin{equation}
J =\frac{2e}{h}\frac{8\pi\Gamma_{L}\Gamma_{R}}{\Gamma}\sin\frac{\phi}{2}\operatorname{Re}\langle d_{\downarrow}d_{\uparrow}\rangle. \label{A10}
\end{equation}

\section{DERIVATION OF the Green Functions in the noninteracting case}
In the noninteracting case ($U=0$), the GFs of the QD are readily obtained through the Dyson equation $G^{-1}(z)=G^{-1}_0(z)-\Sigma(z)$, with $G_0(z)=(zI-H_C)^{-1}$ being the unperturbated GFs of the central region and $\Sigma(z)$ the self-energy resulting from the coupling to the left and right superconducting leads.

Within the Nambu representation \cite{Sun2000,Osawa2008}, under the basis $\Psi=\{d^\dag_\uparrow, d_\downarrow, d^\dag_\downarrow, d_\uparrow, f^\dag, f\}^T$ the GFs of the central region is
\begin{equation}
G(  z)=\langle\langle \Psi, \Psi^\dag\rangle\rangle=\left(
\begin{array}
[c]{cccccc}
\langle \langle d_{\uparrow},d_{\uparrow}^{\dag}\rangle
\rangle  & \langle \langle d_{\uparrow},d_{\downarrow
}\rangle \rangle  & \langle \langle d_{\uparrow
},d_{\downarrow}^{\dag}\rangle \rangle  & \langle \langle
d_{\uparrow},d_{\uparrow}\rangle \rangle  & \langle
\langle d_{\uparrow},f^{\dag}\rangle \rangle  & \langle
\langle d_{\uparrow},f\rangle \rangle \\
\langle \langle d_{\downarrow}^{\dag},d_{\uparrow}^{\dag}\rangle
\rangle  & \langle \langle d_{\downarrow}^{\dag},d_{\downarrow
}\rangle \rangle  & \langle \langle d_{\downarrow}%
^{\dag},d_{\downarrow}^{\dag}\rangle \rangle  & \langle
\langle d_{\downarrow}^{\dag},d_{\uparrow}\rangle \rangle  &
\langle \langle d_{\downarrow}^{\dag},f^{\dag}\rangle \rangle
& \langle \langle d_{\downarrow}^{\dag},f\rangle \rangle \\
\langle \langle d_{\downarrow},d_{\uparrow}^{\dag}\rangle
\rangle  & \langle \langle d_{\downarrow},d_{\downarrow
}\rangle \rangle  & \langle \langle d_{\downarrow
},d_{\downarrow}^{\dag}\rangle \rangle  & \langle \langle
d_{\downarrow},d_{\uparrow}\rangle \rangle  & \langle
\langle d_{\downarrow},f^{\dag}\rangle \rangle  & \langle
\langle d_{\downarrow},f\rangle \rangle \\
\langle \langle d_{\uparrow}^{\dag},d_{\uparrow}^{\dag}\rangle
\rangle  & \langle \langle d_{\uparrow}^{\dag},d_{\downarrow
}\rangle \rangle  & \langle \langle d_{\uparrow}%
^{\dag},d_{\downarrow}^{\dag}\rangle \rangle  & \langle
\langle d_{\uparrow}^{\dag},d_{\uparrow}\rangle \rangle  &
\langle \langle d_{\uparrow}^{\dag},f^{\dag}\rangle \rangle  &
\langle \langle d_{\uparrow}^{\dag},f\rangle \rangle \\
\langle \langle f,d_{\uparrow}^{\dag}\rangle \rangle  &
\langle \langle f,d_{\downarrow}\rangle \rangle  &
\langle \langle f,d_{\downarrow}^{\dag}\rangle \rangle  &
\langle \langle f,d_{\uparrow}\rangle \rangle  &
\langle \langle f,f^{\dag}\rangle \rangle  & \langle
\langle f,f\rangle \rangle \\
\langle \langle f^{\dag},d_{\uparrow}^{\dag}\rangle \rangle  &
\langle \langle f^{\dag},d_{\downarrow}\rangle \rangle  &
\langle \langle f^{\dag},d_{\downarrow}^{\dag}\rangle \rangle
& \langle \langle f^{\dag},d_{\uparrow}\rangle \rangle  &
\langle \langle f^{\dag},f^{\dag}\rangle \rangle  &
\langle \langle f^{\dag},f\rangle \rangle
\end{array}
\right). \label{A10}
\end{equation}
The matrix form of the Dyson equation is
\begin{eqnarray}
&&G^{-1}(  z)=G_{0}^{-1}(  z)  -\Sigma(
z) \notag\\
&&\hspace{-0.4cm}  =\left(
\begin{array}
[c]{cccccc}%
z-\varepsilon_{d\uparrow} & 0 & -\mathcal{F} e^{-i\varphi} & 0 & \lambda &
\lambda\\
0 & z+\varepsilon_{d\downarrow} & 0 & \mathcal{F} e^{-i\varphi} & 0 & 0\\
-\mathcal{F} e^{i\varphi} & 0 & z-\varepsilon_{d\downarrow} & 0 & 0 & 0\\
0 & \mathcal{F} e^{i\varphi} & 0 & z+\varepsilon_{d\uparrow} & -\lambda &
-\lambda\\
\lambda & 0 & 0 & -\lambda & z-\varepsilon_{M} & 0\\
\lambda & 0 & 0 & -\lambda & 0 & z+\varepsilon_{M}%
\end{array}
\right)  -\left(
\begin{array}
[c]{cccccc}
z\Gamma\beta(  z)  &  \Delta \widetilde{\Gamma
}\beta(  z)  &0  &0  &0  &0 \\
 \Delta \widetilde{\Gamma}^{\ast}\beta(  z)  &
z\Gamma\beta(  z)  &0  &0  &0  &0 \\
0 &0  & z\Gamma\beta(  z)  & - \Delta
\widetilde{\Gamma}\beta(  z)  &0  &0 \\
0 &0  & - \Delta \widetilde{\Gamma}^{\ast}\beta(  z)
& z\Gamma\beta(  z)  &0  &0 \\
0 &0  &0  &0  &0  &0 \\
0 &0  &0  &0  &0  &0
\end{array}
\right) \notag\\
&&\hspace{-0.4cm} =\left(
\begin{array}
[c]{cccccc}%
z-\varepsilon_{d\uparrow}-z\Gamma\beta\left(  z\right) & -\Delta\tilde{\Gamma}\beta\left(  z\right)   & -\mathcal{F} e^{-i\varphi} & 0 &\lambda & \lambda\\
-\Delta\tilde{\Gamma}^{\ast}\beta\left(  z\right)   &z+\varepsilon_{d\downarrow}-z\Gamma\beta\left(  z\right) & 0 & \mathcal{F} e^{-i\varphi} & 0 & 0\\
-\mathcal{F} e^{i\varphi} & 0 &z-\varepsilon_{d\downarrow}-z\Gamma\beta\left(  z\right) & \Delta\tilde{\Gamma}\beta\left(  z\right)   & 0 & 0\\
0 & \mathcal{F} e^{i\varphi} & \Delta\tilde{\Gamma}^{\ast}\beta\left(  z\right) &z+\varepsilon_{d\uparrow}-z\Gamma\beta\left(  z\right) & -\lambda & -\lambda\\
\lambda & 0 & 0 & -\lambda & z-\varepsilon_{M} & 0\\
\lambda & 0 & 0 & -\lambda & 0 & z+\varepsilon_{M}%
\end{array}
\right) , \label{A11}
\end{eqnarray}
where $z=\omega\pm i 0^+$ and $\widetilde\Gamma=\sum_l\Gamma_l e^{i(\delta_{lL}-\delta_{lR})\phi/2}$. Obviously, we can readily obtain the GFs of QD by inversing the above matrix.